\documentclass[a4paper,10pt,twoside]{cpc-hepnp}

\usepackage{multicol}
\usepackage{graphicx}
\usepackage{booktabs}
\usepackage{amssymb,bm,mathrsfs,bbm,amscd}
\usepackage[tbtags]{amsmath}
\usepackage{lastpage}
\usepackage{verbatim}

\begin{document}

\fancyhead[c]{\small Chinese Physics C~~~Vol. XX, No. X (201X)
XXXXXX} \fancyfoot[C]{\small 010201-\thepage}

\footnotetext[0]{Received 14 March 2009}

\title{Preliminary study of light yield dependence on LAB liquid scintillator composition
\thanks{Supported by The Strategic Priority Research Program of the Chinese Academy of Sciences£¬Grant No. XDA10010500}
\thanks{Supported by National Natural Science Foundation of China (11390384) }
\thanks{Supported by the CAS Center for Excellence in Particle Physics (CCEPP)}}

\author{%
      YE Xing-Chen$^{1,2;1)}$\email{yexch@ihep.ac.cn}%
\quad YU Bo-Xiang$^{2,3;2)}$\email{yubx@ihep.ac.cn (corresponding author)}%
\quad ZHOU Xiang$^{1;3)}$\email{xiangzhou@whu.edu.cn (corresponding author)}%
\quad ZHAO Li$^{4}$\\
\quad DING Ya-Yun$^{2,3}$
\quad LIU Meng-Chao$^{2,3}$
\quad DING Xue-Feng$^{1}$
\quad ZHANG Xuan$^{2,3}$\\
\quad JIE Quan-Lin$^{1;4)}$\email{qljie@whu.edu.cn (co-corresponding author)}%
\quad ZHOU Li$^{2,3}$
\quad FANG Jian$^{2,3}$
\quad CHEN Hai-Tao$^{2,3}$
\quad HU Wei$^{2,3}$\\
\quad NIU Shun-Li$^{2,3}$
\quad YAN Jia-Qing$^{2,3}$
\quad ZHAO Hang$^{2,3}$
\quad HONG Dao-Jin$^{2,3}$
}
\maketitle

\address{%
$^1$ Hubei Nuclear Solid Physics Key Laboratory, Key Laboratory of Artificial Micro- and Nano-structures of Ministry of Education, and School of Physics and Technology, Wuhan University, Wuhan 430072, China\\
$^2$ State Key Laboratory of Particle Detection and Electronics, Beijing 100049, China\\
$^3$ Institute of High Energy Physics, Chinese Academy of Sciences, Beijing 100049, China\\
$^4$ Nuclear and Radiation Safety Center, Beijing 100049, China\\
}

\begin{abstract}

Liquid scintillator (LS) will be adopted as the detector material in JUNO (Jiangmen Underground Neutrino Observatory). The energy resolution requirement of JUNO is 3 \%, which has never previously been reached. To achieve this energy resolution, the light yield of liquid scintillator is an important factor. PPO (the fluor) and bis-MSB (the wavelength shifter) are the two main materials dissolved in LAB. To study the influence of these two materials on the transmission of scintillation photons in LS, 25 and 12 cm-long quartz vessels were used in a light yield experiment. LS samples with different concentration of PPO and bis-MSB were tested. At these lengths, the light yield growth is not obvious when the concentration of PPO is higher than 4 g/L. The influence from bis-MSB becomes insignificant when its concentration is higher than 8 mg/L. This result could provide some useful suggestions for the JUNO LS.

\end{abstract}



\footnotetext[0]{\hspace*{-3mm}\raisebox{0.3ex}{$\scriptstyle\copyright$}2013
Chinese Physical Society and the Institute of High Energy Physics
of the Chinese Academy of Sciences and the Institute
of Modern Physics of the Chinese Academy of Sciences and IOP Publishing Ltd}%

\begin{multicols}{2}

\section{Introduction}

Liquid scintillator (LS), for its low price and high light yield, is widely used in the neutrino detection. It has performed well in experiments including KamLAND\cite{lab1}, CHOOZ\cite{lab2}  and Daya Bay\cite{lab3}.

Liquid scintillator can be a mixture of a solvent, a little fluor and very low concentration of wavelength shifter. For example, the liquid scintillator used in the Daya Bay Neutrino Experiment consists of linear alkyl benzene (LAB) as the solvent, 3 g/L 2,5-diphenyloxazole (PPO) as the fluor, and 15 mg/L p-bis-(o-methylstyryl)-benzene (bis-MSB) as the wavelength shifter.\cite{lab4}

In a LS detector, a reactor neutrino signal begins with an inverse ¦Â-decay reaction, emitting a positron. Then the positron annihilates with an electron into two 0.511 MeV ¦Ãs. The ¦Ãs transfer energy to two electrons through Compton scattering. The Compton electrons deposited energy in the liquid scintillator via ionization. The deposit energy is then converted into scintillation photons and collected by photomultiplier tubes (PMTs). In a LS detector, a series of ¦Ãs with a certain energy can finally form scintillation photons with a certain energy spectrum. That is the light yield of the liquid scintillator.\cite{lab5}

Light yield is very important for scintillator detectors, especially when the scintillation photons have to travel a long distance before they reach the PMTs. A higher light yield means more information from an event. For the JUNO (Jiangmen Underground Neutrino Observatory) experiment, which plans to use a 35-meter-diameter spherical neutrino detector, a liquid scintillator with high light yield is therefore being searched for.

In this study of propagation of scintillation photons, the transmission spectrum and light yield of LS is investigated with different compositions of LS. Section 2 presents experimental setup and results for measurement of the transmission spectrum, Section 3 presents the study of relative light yield, and Section 4 draws some conclusions.

\section{LS Transmission Spectrum }

\subsection{Experimental Setup}

\begin{center}
\includegraphics[width=4cm]{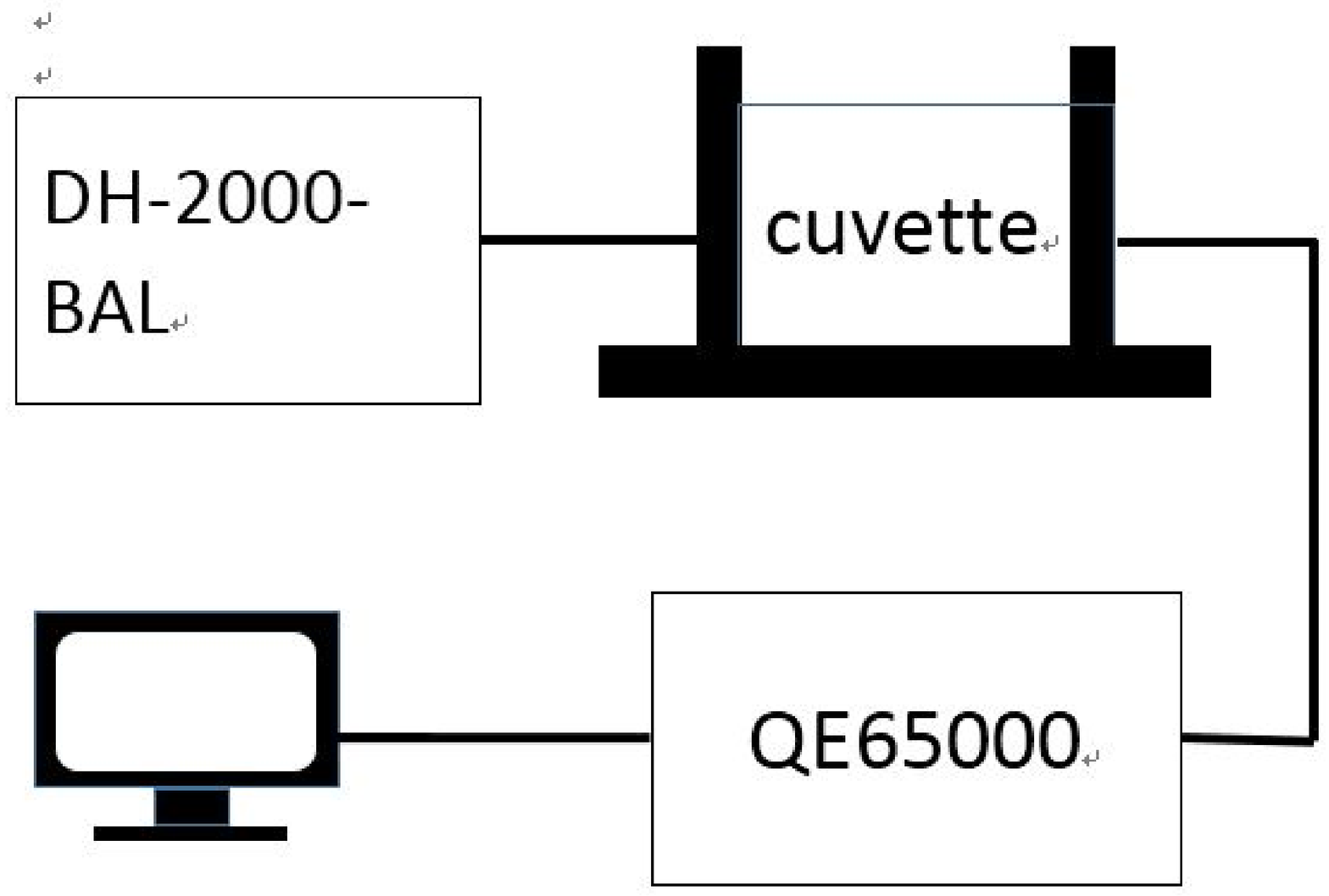}
\figcaption{\label{fig1}   Transmission Spectrum Measurement Setup  }
\end{center}

The principle of this experiment is to measure the transmission spectrum of LS in cuvettes of different lengths. LS of various concentrations of PPO and bis-MSB were tested. In this experiment, seven cuvettes were used. The lengths of light path were 1 cm, 2 cm, 4 cm, 6 cm, 7 cm, 8 cm, and 9 cm.

The transmission spectrum measurement system is composed of three parts: a light source with a continuous spectrum, a cuvette in a cassette, and a spectrometer, as shown in Fig.1. The light source is a Mikropark DH-2000-BAL. There are two lamps emitting light of respectively shorter  and longer wavelength. The lamp used in this experiment has a spectrum from 200 nm to 450 nm. The cassette is made of aluminum frame and covered with black paper. The spectrometer is an Ocean Optics QE65000.\cite{lab6}

\subsection{Experimental Results}

In this experiment, the lamp emits a continuous spectrum from 200 nm to 450 nm. After the light has travelled through some thickness of LS, the spectrum of shorter wavelength is absorbed, and an absorption edge appears. ( Fig. 2) As the cuvette gets longer, a trend can be seen that the absorption edge goes a longer wavelength while the spectrum strength of the same wavelength goes down. That means light of that wave length is strongly attenuated in the LS. When the cuvette is longer than a certain length, however the absorption edge stops at a certain position. That means when light travels further than that thickness in the LS, the light, the wavelength of which is shorter than that of the absorption edge, has largely attenuated.

Three samples of LS with PPO (1.5 g/L, 3 g/L and 5 g/L ) and three samples with PPO ( 3 g/L ) and bis-MSB ( 5 mg/L, 15 mg/L and 30 mg/L ) were tested. The experiment data of samples with 3 g/L PPO and with 3 g/L PPO and 15 mg/L bis-MSB are shown in Fig.  2 and Fig.  3 respectively. The horizontal axis shows the wave length. The vertical axis shows the relative spectrum strength. In each figure, 7 curves are shown, representing the transmission spectrum for different lengths of LS. Though only two figures are shown here, the results of the other samples are relatively similar. When the cuvette is longer than 6 cm, the transmission spectrum changes only a little, no matter which sample is measured. The absorption edge stops at 6 cm, and we are interested in the light yield farther than that distance in LS. So two quartz vessels longer than that length were made for the light yield experiment.

\begin{center}
\includegraphics[width=9cm]{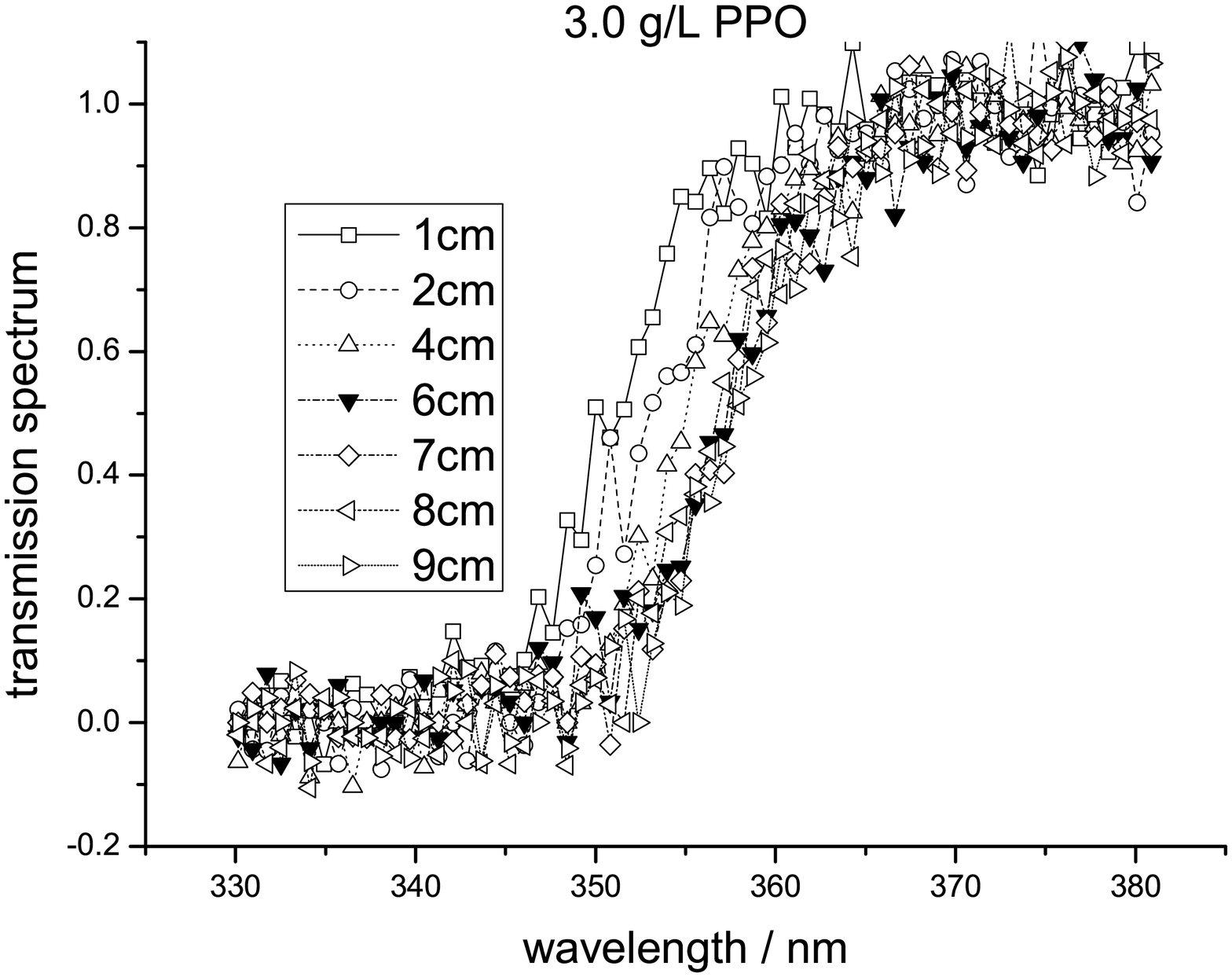}
\figcaption{\label{fig3(b)}    Transmission spectrum of LS with 3 g/L PPO  }
\end{center}

\begin{center}
\includegraphics[width=9cm]{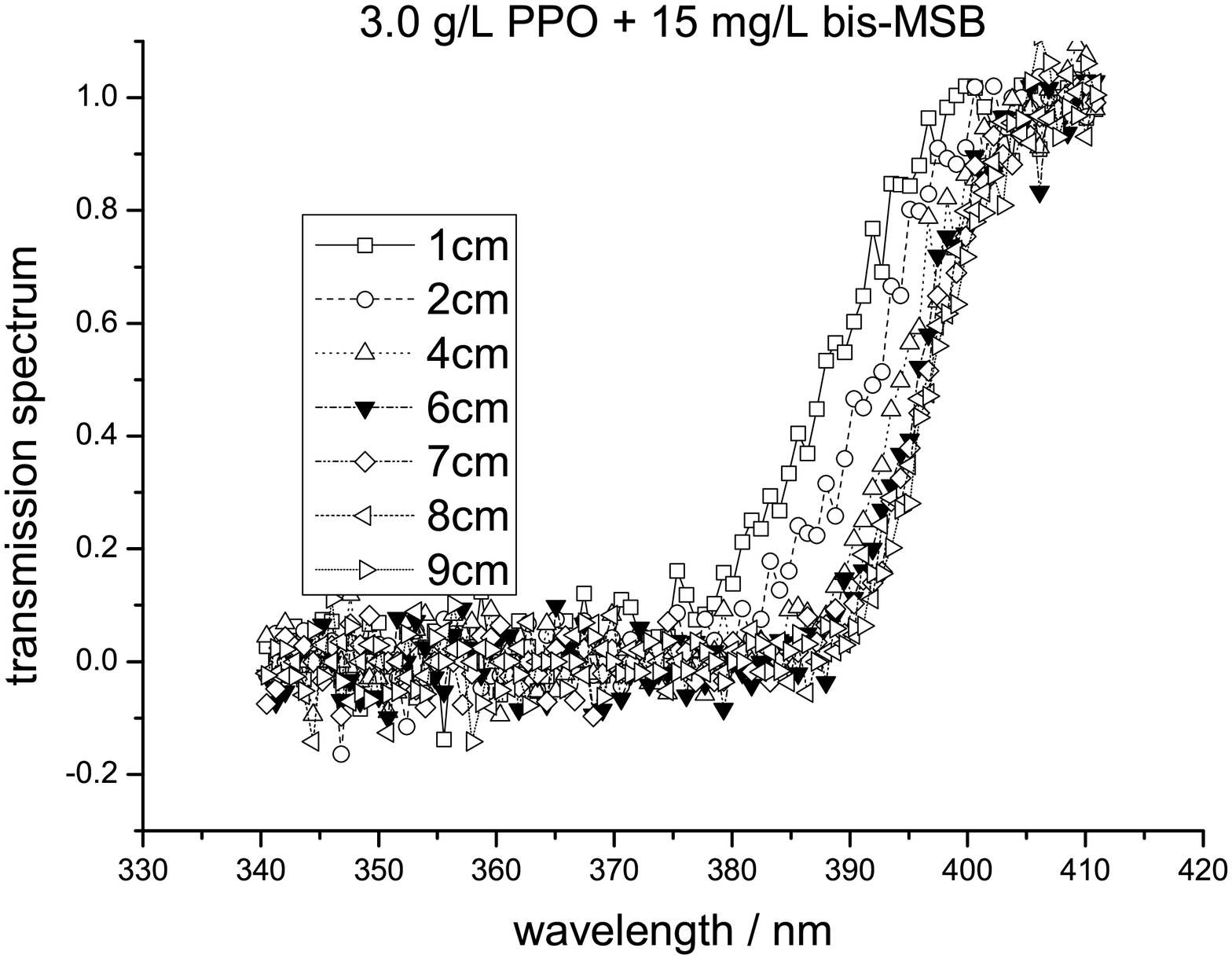}
\figcaption{\label{fig3}  Transmission spectrum of LS with 3 g/L PPO and 15 mg/L bis-MSB   }
\end{center}

\section{The Relative Light Yield of LS}

\subsection{Experimental Setup}

\begin{center}
\includegraphics[width=7cm]{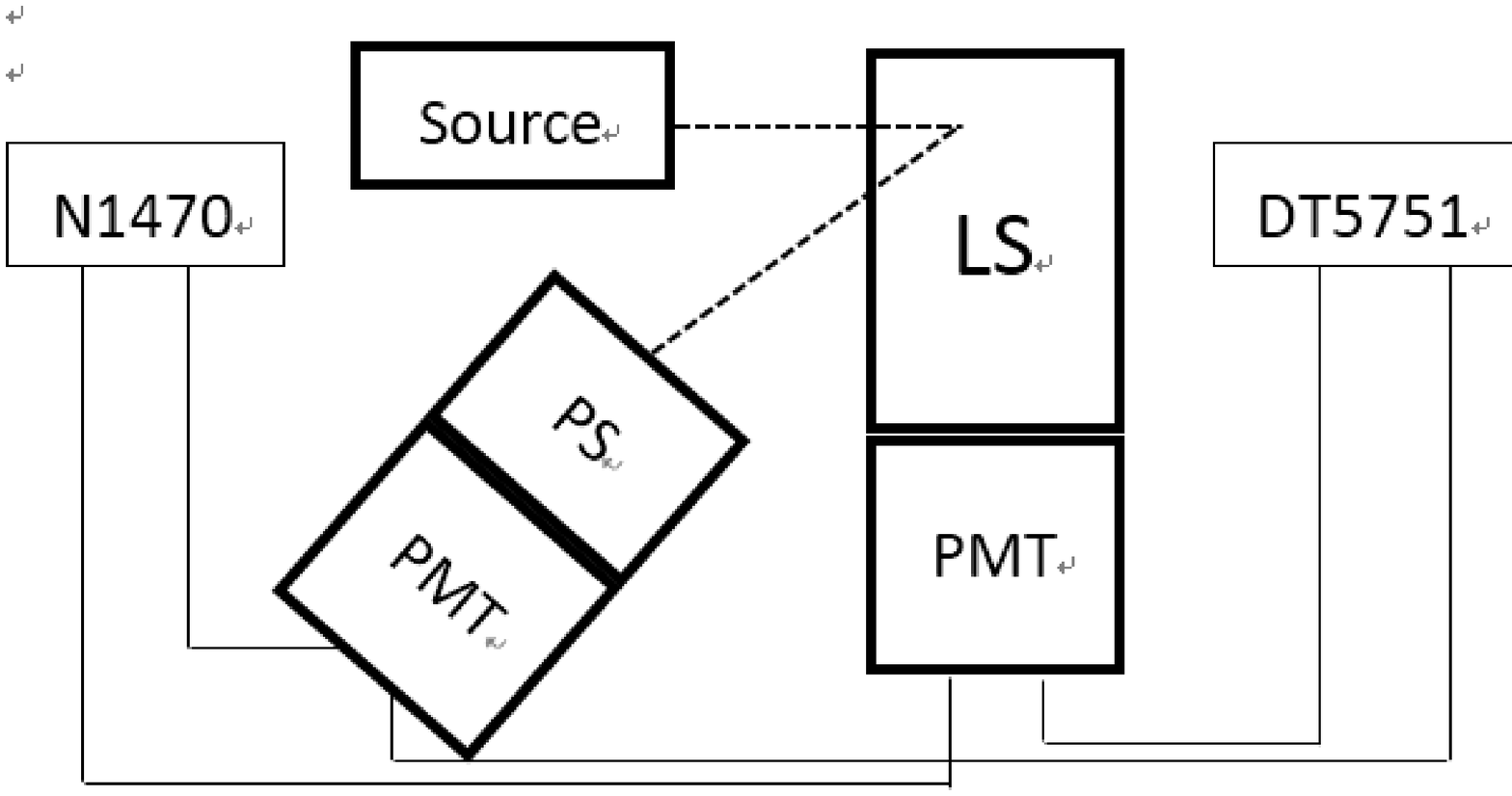}
\figcaption{\label{fig4}   Setup of Light Yield Measurement  }
\end{center}

The experimental setup of light yield measurement is shown in Fig.  4. The radiation source emits a gamma photon, which scatters with an electron in the LS. The gamma photon is then scattered into another direction at random. When the coincidence detector set at a certain direction is hit by the photon, it is triggered and sends out a trigger signal, so the recoil electron events with a relative narrow energy distribution can be picked out, as shown in Fig.space5. The signal is recorded by the computer after digitalization, and further processed by LabView. Two quartz vessels, 25 cm and 12 cm in height, were used in this experiment as LS containers. The gamma photons hit near the top of the vessels while the PMT is set at the bottom. The type of PMT is xp2020. The digitizer is CAEN DT5751.\cite{lab7}

\begin{center}
\includegraphics[width=8cm]{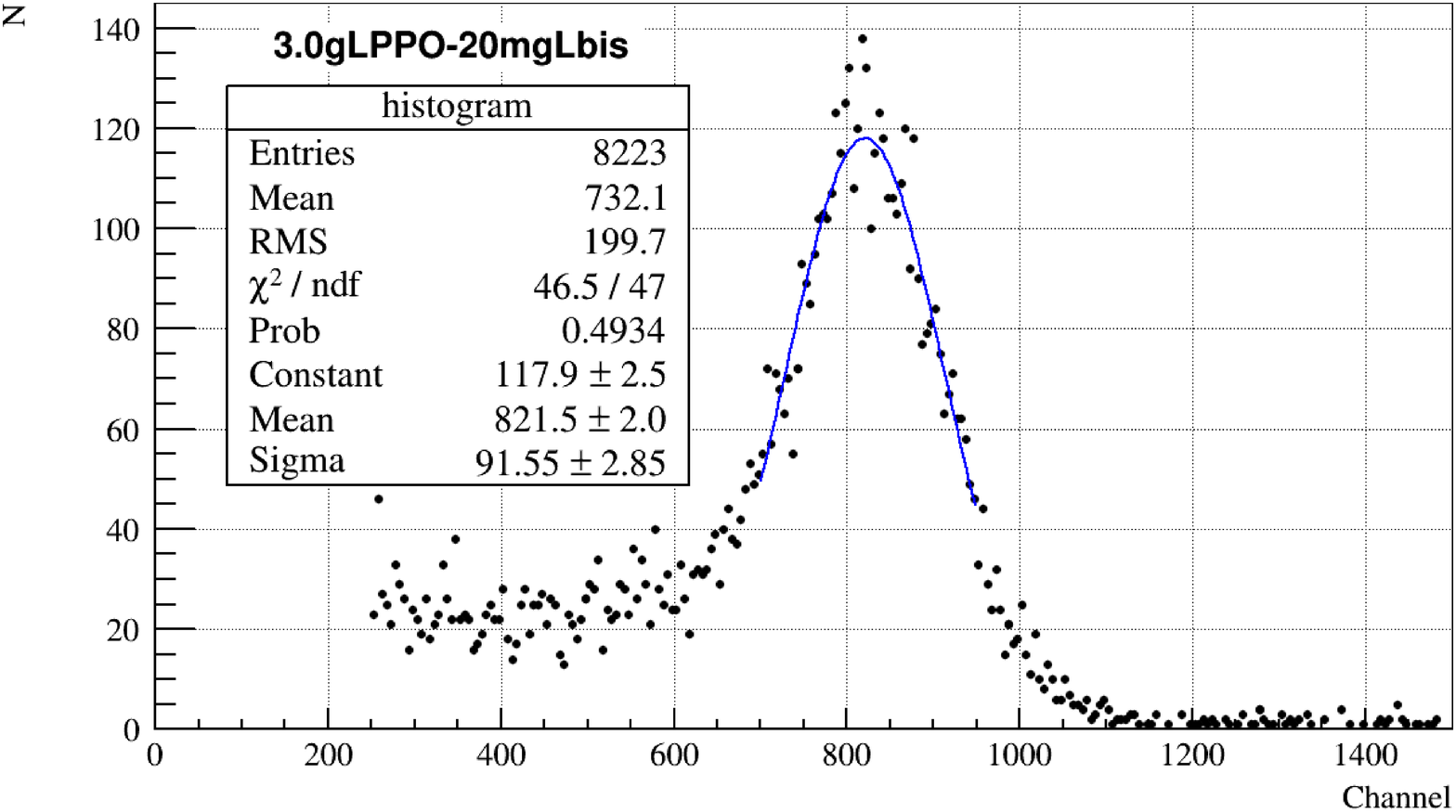}
\figcaption{\label{fig5}   Light Yield Spectrum }
\end{center}

Fig. 5 shows the light yield spectrum. The vertical axis is the scintillation events count, and the horizontal axis is the channel. The relative light yield of LS is the ratio of the peak position. The peak was fitted with a Gaussian function. From each sample we got a peak position.

\begin{center}
\includegraphics[width=9cm]{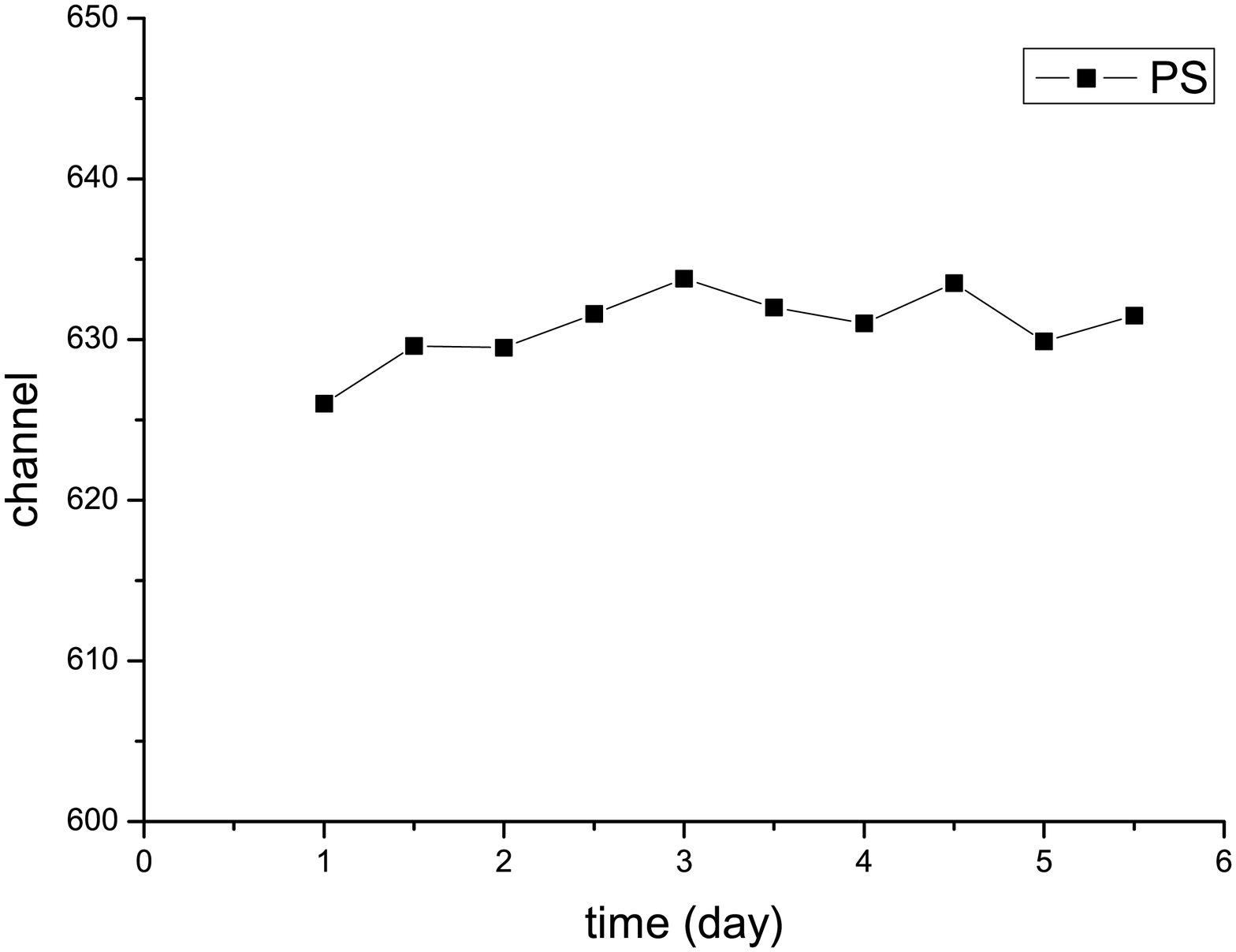}
\figcaption{\label{fig6}   System Stability Measurement  }
\end{center}

Fig. 6 shows the stability of the light yield measurement system. We used the same plastic scintillator to monitor the measurement system. Each point in the curve is the relative light yield of that plastic scintillator, tested before and after testing a set of LS.

\subsection{Experimental Results}

The light yield of 45 LS samples were measured. The concentrations of PPO were 1.0, 1.5, 2.0, 2.5, 3.0, 3.5, 4.0, 5.0, and 7.0 g/L. Each concentration of PPO corresponded to 5 concentrations of bis-MSB, 5, 8, 12, 15, 20 mg/L.

The samples were measured twice in different orders. According to the concentration of bis-MSB, the samples were first divided into 5 sets, within which the concentration of PPO rises. All samples were measured in a 25-cm-long quartz vessel. A plastic scintillator was used to calibrate the whole measurement system. The plastic scintillator was measured before and after the measurement of each set. During the whole period, the electronics was kept powered on. The temperature was kept around 23 ¡æ, with a maximum of 23.06 ¡æ and a minimum of 22.68 ¡æ, so the influence from the temperature distribution can be ignored.

The samples were then measured in another order. They were divided into 9 sets according to the same concentration of PPO. Within each set, the concentration of bis-MSB rises. All sample were measured in a 12-cm-long quartz vessel. A plastic scintillator acted as the calibrator. The temperature was kept around 20 ¡æ, with a maximum of 20.06 ¡æ and a minimum of 19.50 ¡æ.

Fig. 7 shows the results of the measurement in the first order. The horizontal axis is the concentration of PPO, and the vertical axis is the relative light yield. All data was normalized according to Daya Bay LS (3 g/L PPO and 15 mg/L bis-MSB). Five curves are shown in the plot. Each curve represents the light yield of LS with the same concentration of bis-MSB.

\begin{center}
\includegraphics[width=9cm]{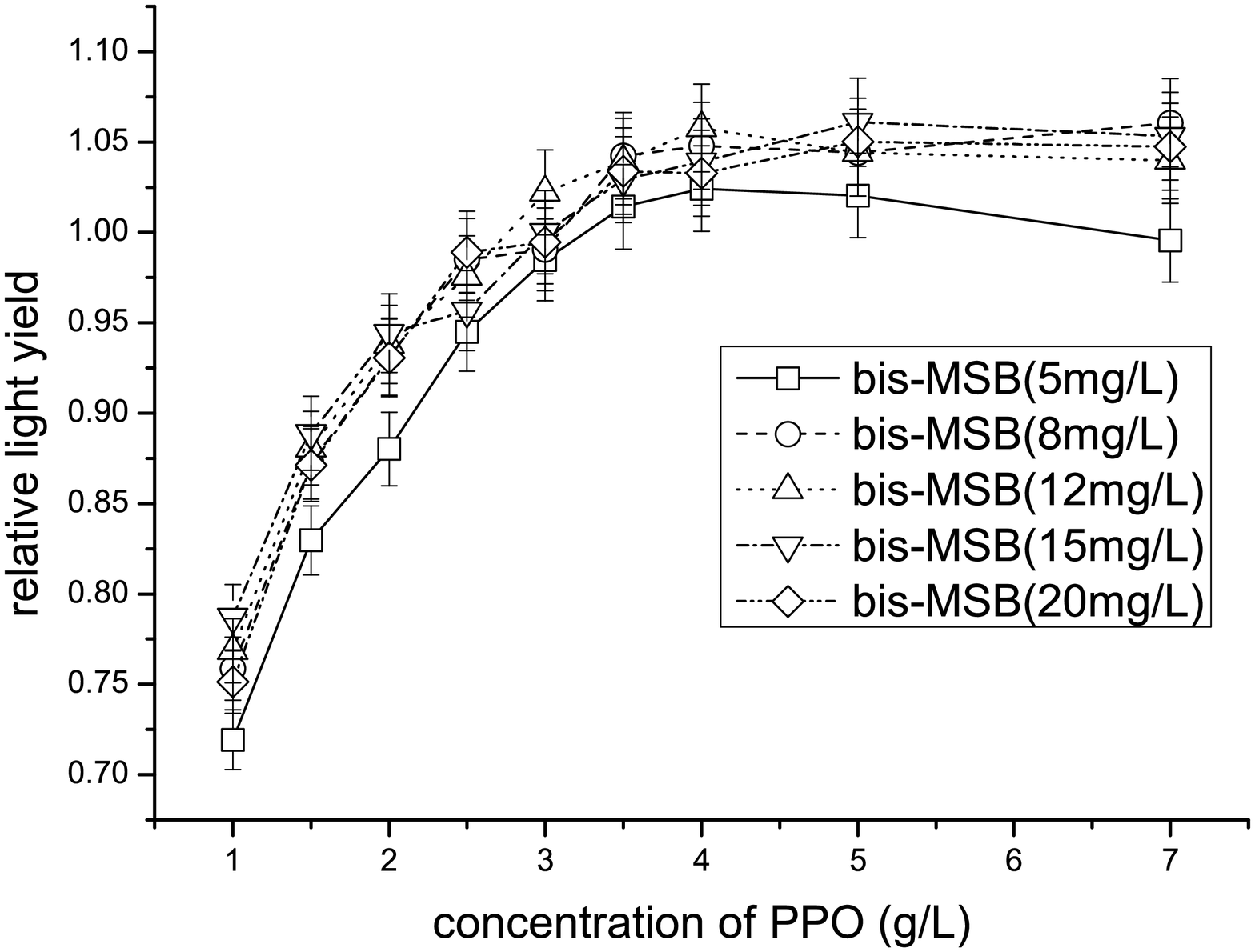}
\figcaption{\label{fig7}  Relative Light Yield Result according to the Concentration of PPO  }
\end{center}

\begin{center}
\includegraphics[width=8cm]{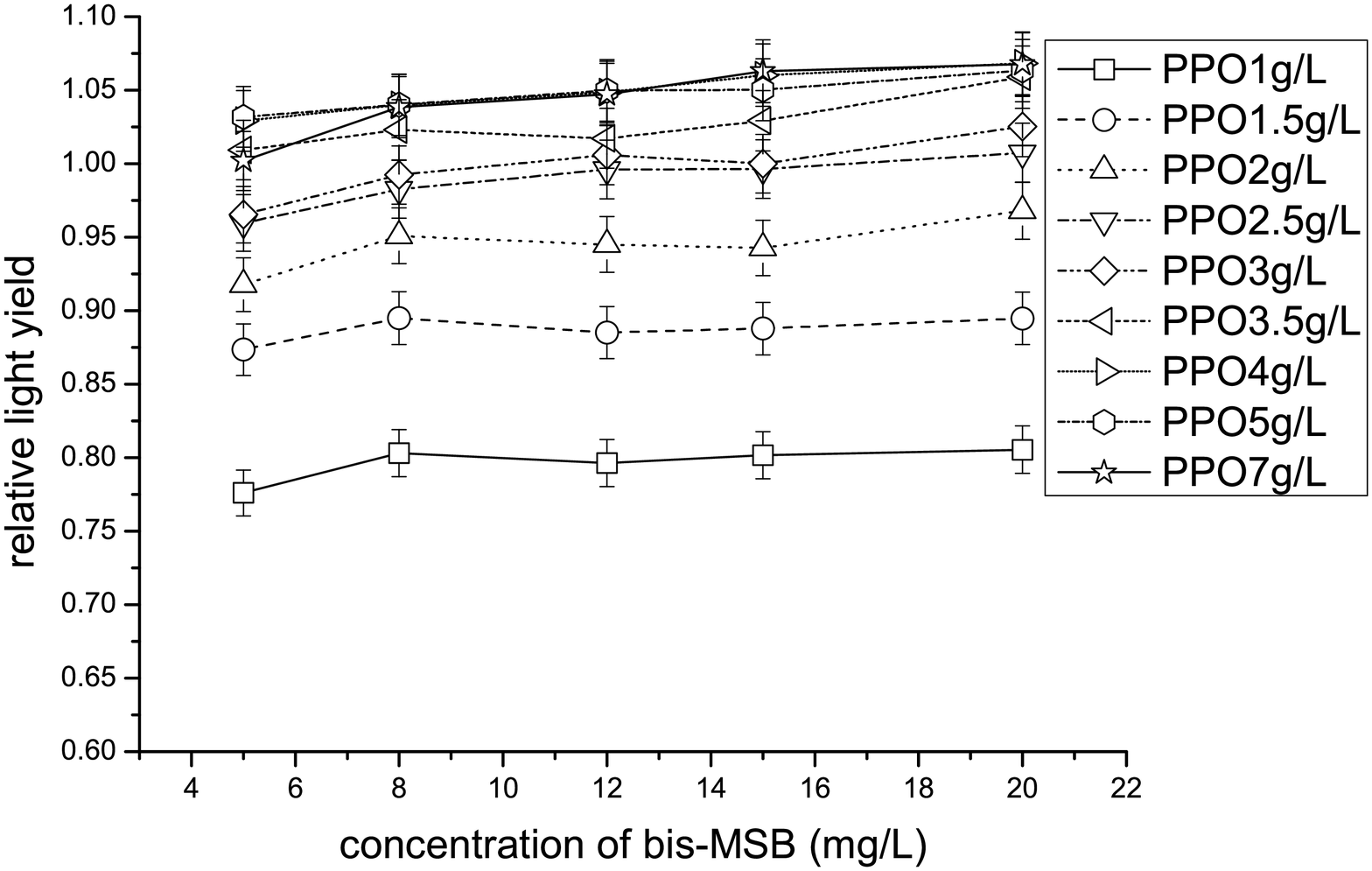}
\figcaption{\label{fig8}  Relative Light Yield Result according to the Concentration of bis-MSB  }
\end{center}

Fig. 8 shows the results for the second order. The horizontal axis is the concentration of bis-MSB. Nine curves are shown in that plot, one for each set of measurements.

\begin{center}
\includegraphics[width=9cm]{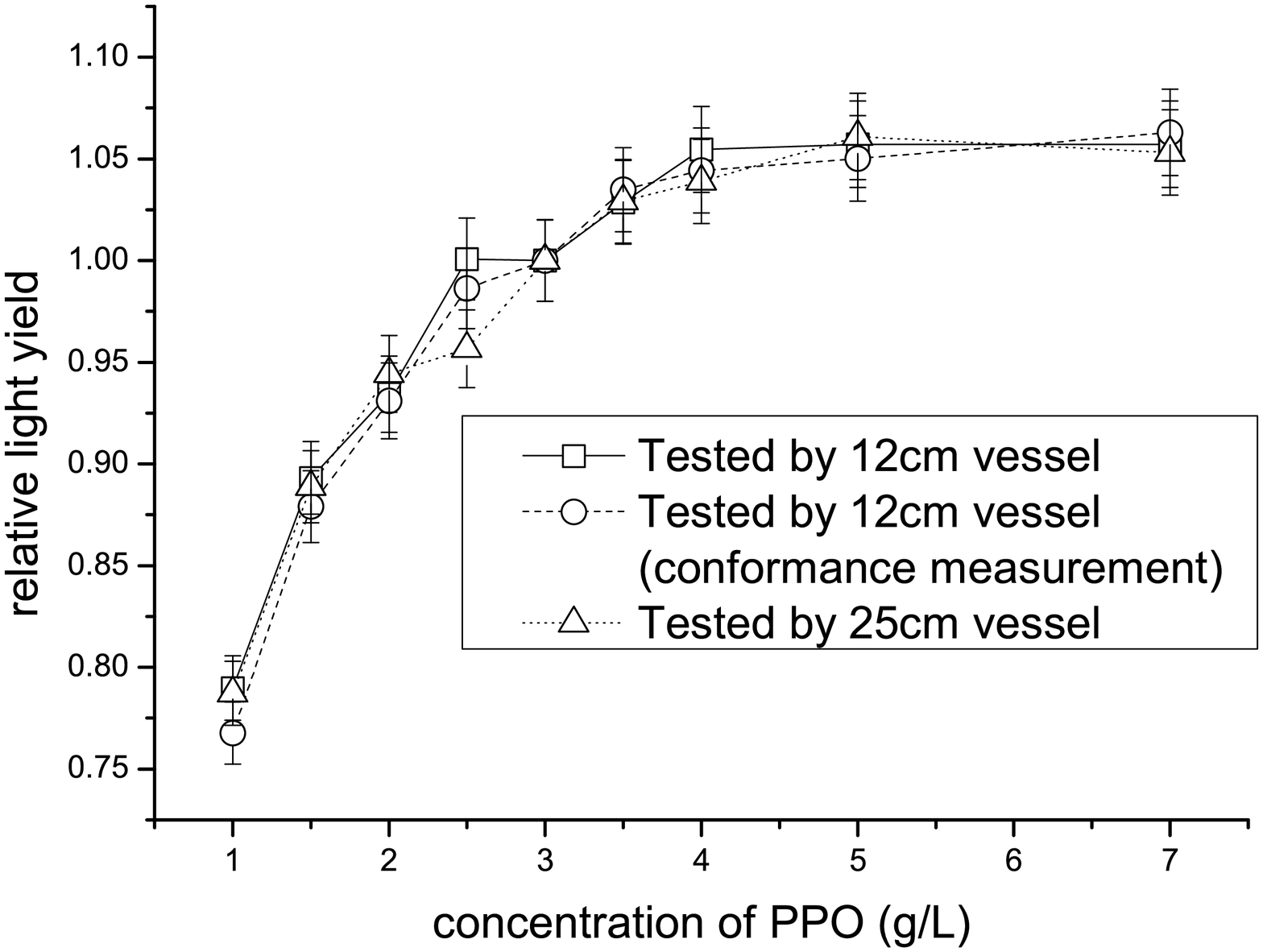}
\figcaption{\label{fig9}  Results Comparison  }
\end{center}

In order to ensure the consistency of the results measured with the two vessels, data from the two measurement orders were then compared. Three curves are shown in Fig. 9. All of them are from the same set of samples, with fixed concentration of bis-MSB, 15mg/L. The concentration of PPO rises from 1 g/L to 7 g/L. The squares show the data picked out from each set in the second measurement order. The circles are the data measured independently with the 12-cm-long quartz vessel. The triangles show one of the sets from the first measurements order. From this figure, it can be seen that measurement of the two orders match well.

Measurements of the two orders both lead to the conclusion that light yield growth is not obvious when the concentration of PPO is higher than 4 g/L. In this experiment, the light yield growth as a function of concentration of bis-MSB becomes insignificant after 8 mg/L.

\end{multicols}
\begin{multicols}{2}

\section{Summary}

Through measurement of the scintillation photon transmission spectrum, it was found that the transmission spectrum changes very little after passing the 6 cm mark in LS.

The light yield, measured with vessels of 20 cm and 10 cm in length, shows no significant growth when the concentration of PPO is higher than 4 g/L. Compared with the concentration of Daya Bay LS, 3 g/L PPO and 15 mg/L bis-MSB, concentration of PPO could be raised a little to 4.0 g/L. The influence from the concentration of bis-MSB on the light yield, however, seems not to be significant when it is higher than 8 mg/L. To understand how bis-MSB influence the light yield in a larger detector, further research will be done in Daya Bay AD detector.

\end{multicols}
\vspace{-1mm}
\centerline{\rule{80mm}{0.1pt}}
\vspace{2mm}
\begin{multicols}{2}

\end{multicols}

\clearpage

\end{document}